\documentstyle[epsfig,12pt]{article}

\begin{document}

\begin{center} 
{\Large NONPERIODIC OSCILLATIONS OF PRESSURE
IN \\~ \\
 A SPARK IGNITION COMBUSTION ENGINE
} \\~ \\ ~ \\
M. WENDEKER, G. LITAK, J. CZARNIGOWSKI AND K. SZABELSKI \\~
\\ Department of Mechanics, Technical University of Lublin,
Nabystrzycka 36,
20-618 Lublin, Poland 

\end{center}


\begin{abstract} 
We report our results on non-periodic
experimental time series of pressure in a spark ignition engine.
The experiments were performed for a low rotational velocity of
a crankshaft and a relatively large spark advance angle. We show that
the combustion process has many chaotic features. Surprisingly,
the reconstructed attractor has a characteristic butterfly shape
similar to a chaotic attractor of Lorentz type.
The suitable recurrence plot shows that the dynamics of the combustion is
a nonlinear multidimensional process mediated by stochastic noise.
\end{abstract}

\section{Introduction} 
The cycle-to-cycle combustion variability
has been a subject of interest for many years [Heywood 1988, Hu 1996].
Noise in pressure time dependence is a factor
which makes an engine control difficult [Wendeker \&  Czarnigowski 2000].
To make the progress some papers concentrated on
stochastic models formulation [Roberts {\em et al.} 1997,  Wendeker {\em et al.} 1999]. However,
as identified by Heywood [1988], the sources of fluctuations involve various
process conditions such as: aerodynamics in the cylinder during combustion,
the amount of fuel, air and recycled exhaust gases supplied
to the cylinder and a mixture composition near the spark plug.
It is possible that many different disturbances influence the
process making it stochastic
but one should note that high sensitivity  on the process conditions is
characteristic for
nonlinear phenomena and deterministic chaos.  In that case the pressure
variations might originate from
a complex dynamics leading, presumably with some stochastic component,
to nonperiodic behaviour [Daw {\em et al.} 1996 and 1998,  Wendeker 2002]. Thus, the crusial problem  is 
to understand
the nonlinear dynamics of the process observing the internal pressure
inside the cylinder [Antoni {\em et al.} 2002]. In the present note we discuss experimental
results of a direct measurement of pressure [Czarnigowski 2002].
Namely, we start our analysis with study of the measured  time series of pressure.
The crankshaft frequency was stabilized by a  electromechanical brake to
the value $\Theta$ =10Hz ($\Theta$=600 RPM). Note, in a  combustion process one of each
two cycles of a crankshaft rotation are  coincided with combustion leading
to pressure parametrically driven  oscillation with a period $T=2/\Theta=0.2$s.
In Fig. 1 we  show a typical time history of 
combustion internal pressure for  $\Delta \alpha_z =\pi/6$ ($\Delta \alpha_z =30^o$). We took  
such process conditions 
(a low rotational velocity $\Theta$ and a large advance angle $\Delta \alpha_z$) purposely to make  
nonperiodicity of time
history more transparent [Wendeker {\em et al.} 2002]. Our task in the next  sections will be to
examine it  by using tools for nonlinear analysis.

\section{Nonlinear Analysis}

Our first step is reconstruction of the dynamic attractor.  Having time
series limited only to 200 cycles it is not possible to  perform full
analysis of embedding and correlation dimension of the  dynamical process
[Kantz \& Scheiber 1997, Abarbanel 1996] so we decided to do a simplified study. In Fig.  2 we show a 
two dimensional attractor obtained using a time delay procedure  introduced by
Takens [Takens 1981]. Thus, we introduce a new vector {\bf p}:
\begin{equation}
{\bf p}=[ p(t), p(t-\Delta t), p(t- 2\Delta t), ...]
\end{equation}
but here we will use only the minimal set of variables: p(t)  and p(t-(t).
The time delay $\Delta t=0.977 T$ is slightly smaller than the  driving period
$T=2/\Theta$. Surprisingly, the reconstructed attractor (Fig. 2) has  a characteristic
butterfly shape similar to a chaotic attractor of a Lorentz system [Lorentz 1963].
Realy, it shows some deterministic  characteristics. For
instance, it consists of two well defined closed  lops. The motion along
first and second loops with possible random switch between them creates
a new question how much of determinism is included in prsesure dynamics. In aim to solve
this problem we decided to examine recurrence plots.  This method,
originally introduced by Eckmann {\em et al.} [1987],  is capable to
distinguish deterministic system from stochastic one.  It has been already
applied successfully to various nonlinear systems [Casdagli 1997,  Kantz \& Scheiber 1997].
In order to make a recurrence plot for trajectory on  reconstructed phase
space (Fig. 2) we calculate the distance between each  pair of measured ${\bf p}_i
=[p(t_i),p(t_i - \Delta t)]$ , where $t_i$ is a sampling time.  Following Eckmann {\em et al.} [1987] we 
define a 
new
matrix {\bf R} with elements:
\begin{equation}
 R_{ij} \equiv {\rm H}(\epsilon - |{\bf p}_i -{\bf p}_j|),  
\end{equation}
where ${\rm H}(x)$ denotes the Heaviside step function and $\epsilon$ is a small positive number. 
Its value ($\epsilon$) is usually assumed to be a few percents of the maximal distance between ${\bf p}_i$ 
and ${\bf p}_j$. 
{\bf R} is a square matrix, which has 
the size $M \times M$, where $M$ is a number of measured points in the collected experimental data. After
defining ${\bf R}$ we plot black point in $i,~j$ plane if only $R_{ij} =1$. In this way black regions can 
indicate close
returns   of  trajectories or some flat regions is time dependence when a recurrence plot  correspond to
shorter intervals. The suitable plots for studied time history were plotted in Fig. 3a and b.

On the other hand, Fig. 3c  shows the recurrence plot for series of numbers obtained from a random
numbers generator with a  uniform distribution. To be more specific, in Fig. 3a, we show a recurrence
plot for 10 sequential cycles  of the engine work. The regular patterns identify deterministic flow in
this scale. Note however, that  the unstable region of a pressure fluctuation
 corresponding to white regions 
is connected only with
some specific time interval of combustion.  It is clearly depicted in Fig. 4a where we plotted following
stroke phases of each combustion cycle: intake of gases,  compression, power creation and finally exhaust of
gases and combustion products. The recurrence plot  in Fig. 3b was plotted for a larger scale of 200
cycles. In this case we used only chosen experimental  data pints for a given crank angle $\phi=2.125\pi$. In
this new scale the plot is not so much regular as Fig.  3a but still differs from the case of random
numbers (Fig. 3c). First of all it possesses patterns of  vertical and horizontal black stripes. Such
structure can be identified as belonging to chaotic behaviour.  Note, that size of ${\bf R}$ matrix (Eq. 2) 
was
limited, by the available short time series, to only $200 \times 200$.  Obviously, in larger time scale we 
expect
black stripes to be represented by corresponding vertical and horizontal  lines. Another problem is
connected with stochastic noise which may disturb the time correlations in  that picture. In Fig. 4a we
show also three time histories selected from the studied experimental time  series of pressure (Fig. 1).
One can easily note that curves '1' and '2' correspond to combustion with  different, rather high,
efficiency while the curve '3' is a case where the combustion is very small  or even not present at all.
This effect around a top point of compression (top dead center) for a crank angle $\phi=2\pi$ is even  better 
visible. In this purpose we magnified that region in Fig. 4b.
Clearly, one can see the difference of pressure  between $\phi=2\pi \pm 0.125 \pi$
(note 'A' and 'B' dashed lines in Fig. 4b). For better clarity, we have also plotted  (Fig. 4c) the time
series of points $p_n$ connected with these two different phases ('A' and 'B') of the engine  work. Only in 
case
of combustion (curve 'A' in Fig. 4c) we see the fluctuation of pressure. Really, this  justify of our
choice of combustion sensitive dada to a recurrence plot in Fig. 3b. The lack of clear white  regions  in 
Fig. 3b
can be the result of higher (than 2) embeding dimension of the dynamical system and the  output signal
modulation visible in Fig. 4c.

\section{Intermittency as a Route to Chaos in a Combustion Process}

Both, the curve 'A' in Fig. 4c and the previously presented dynamical attractor (Fig. 2) show some features  
typical for  
systems showing intermittent chaos [Pomeau \& Manneville 1980, Chatterjee \&  Malik 1996]. To clarify this 
point we plotted once again a  time history
of a studied case for a shorter time interval (few cycles) in Fig. 5a. Note that after a sequence  of a
steady flow (with very weak combustion) in cycles marked as 1, 2, 3 we have sudden jump of pressure $p$ in 
the next cycle  (marked by 4
in Fig. 5a). In Fig. 5b we have resolved this behaviour plotting trajectory of system for  these 4   
cycles. Interestingly the sudden jump of pressure can be associated with escape from the  present
trajectory. Such jump is a characteristic phenomenon for an intermittency mechanism of a  chaotic motion.
Note, the above behaviour is typical for that system as trajectory presented in Fig. 5a.  is a  
part 
of that plotted in Fig. 2. Our system is already in a chaotic state so we can find only some short 
intervals which behave in this way.
Interestingly, these jumps of high  efficient combustion follows the sequence of cycles with weak
combustion (Fig. 5a, Fig. 4c).  Physically, this means that it happens if the amount of residual exhaust
gases from preceding combustion  cycles are absent in the cylinder and, simultaneously, the present
mixture of fuel and air amounts is rich. Such explanation is consistent with the simple nonlinear model
derived by Daw {\em et al.} [1996 and 1998]. According  to their finding the residual gases alter the 
in-cylinder
fuel-air ratio and finally the combustion efficiency in succeeding cycles.

\section{Conclusions and Last Remarks}

Summering our results we would like to add  that we examined the oscillations of internal combustion
pressure in an engine cylinder. It has appeared  that, in some conditions, intermittency is capable to
drive the system into a chaotic region. The strange  attractor presented in Fig. 2 (and Fig. 5b) has many
features typical for systems showing intermittent chaos.  Moreover the recurrence plots indicate that the
system is not likely to be stochastic, but we cannot exclude some influence of random disturbances.
Our results throw some new light on the problem of noise  present in combustion engines and can also give a
new concept to fight with it. However to be more sure about  the physical situation one should perform
studies of longer time series. This can enable us to use other  more sophisticated techniques of nonlinear
dynamics analysis [Kantz \& Scheiber 1997,  Abarbanel 1996]. Such work is in progress and results will 
be reported in a separate publication


\section*{References}

~

Abarbanel, H.D.I. [1996] {\em Analysis of Observed Chaotic Data} (Springer-Verlag, New York).

Antoni, I., Daniere, J., Guillet, F. [2002] "Effective vibration analysis of ic engines  
using 
cyclostationaryrity.
part II - new results on the reconstruction of the cylinder pressures", {\em J. Sound 
Vibr.} 
{\bf 257}, 839-856.

Casdagli, M.C. [1997] "Recurrence plots revisited", {\em Physica} {\bf D 108}, 12-44.

Chatterjee, S., Malik, A.K. [1996] "Three Kinds of Intermittency in a Nonlinear
System", {\em Phys. Rev.} {\bf E 53}, 4362-4367. 

Czarnigowski J. [2002]  unpublished, {\em PhD Thesis, Technical University of Lublin}. 

Daw, C.S., Finney, C.E.A., Green Jr., J.B., Kennel, M.B., Thomas, J.F. and Connolly, F.T.  [1996] "A simple 
model 
for cyclic
variations in a spark-ignition engine", {\em SAE Paper} 962086.

Daw, C.S., Kennel, M.B., Finney, C.E.A., Connolly, F.T. [1998] "Observing and modeling  nonlinear dynamics in 
an
internal combustion engine", {\em Phys. Rev.} {\bf E 57}, 2811-2819.

Eckmann, J.-P., Kamphorst, S.O. and Ruelle, D. [1987] "Recurrence Plots of 
 Dynamical Systems", {\em Europhysics Letters} {\bf 5}, 973-977.

Heywood, J.B. [1988] {\em Internal Combustion Engine Fundamentals} (McGraw-Hill, New 
York).

Hu, Z. [1996] "Nonliner instabilities of combustion processes and 
cycle-to-cycle variations in spark-ignition engines", {\em SAE Paper} 961197.

Kantz, H., Scheiber, T. [1997] {\em Nonlinear Time Series Analysis} (Cambridge University Press, Cambridge). 

Lorentz, E.N. [1963] "Deterministic nonperiodic flow", {\em J. Atmos. Sci.} {\bf 20}, 130.

Pomeau, Y., Manneville, P. [1980] "Intermittent transition to turbulence in
dissipative systems" {\em Commun. Math. Phys.} {\bf 74}, 189.

Roberts, J.B., Peyton-Jones, J.C., Landsborough, K.J. [1997] 
"Cylinder Pressure Variations as a Stochastic Process", {\em SAE Paper} 970059.

Takens, F. [1981] {\em Lecture Notes in Mathematics}  {\bf 898} (Springer, Heidelberg).

Wendeker, M., Niewczas, A., Hawryluk, B. [1999] "A stochastic model of the 
fuel injection of the si engine", {\em SAE Paper} 00P-172.

Wendeker, M., Czarnigowski, J. [2000]  "Hybrid air/fuel control using the adaptive 
estimation 
and neural network", {\em SAE Paper} 2000-01-1248.

Wendeker, M., Czarnigowski, J., Litak, G., Szabelski, K. [2002] "Chaotic combustion in 
spark ignition engines", 
preprint nlin.CD/0212050.


\newpage

\section*{Figure Captions}
\begin{enumerate}
\item{Figure 1.}  Experimental time series of pressure $p(N)$ in a combustion process for
a crankshaft frequency $\Theta=600$ RPM and  a spark advance  angle $\Delta \alpha_z=\pi/6$.   
Note, time is represented by sequential cycles $N$.
\item{Figure 2.} A 'butterfly' attractor reconstructed from  the experimental  
time series of pressure. $\Delta t=0.977 T$.
\item{Figure 3.}  Recurrence plots of the examined combustion process: in a small scale of 10 cycles 
(Fig. 
3a),
and the scale  of 200 cycles for selected points chosen for the same crank angle $\Theta=2.125\pi$ (Fig. 3b). 
Fig.
3c shows the  recurrence plot for the series of numbers from random numbers generator with a uniform
distribution.
\item{Figure 4.}  Time history  of combustion presented in Fig. 1 plotted for three chosen cycles (Fig. 
4a) 
and
magnification of it  (Fig. 4b) around the top dead point (for a crank angle $\phi=2\pi$). Fig. 4c shows a 
sequence of 
pressure
values for given crank angle $\phi=2\pi \mp 0.125 \pi$ (for B and A as before and after ignition, 
respectively).
\item{Figure 5.}
Interval of the time history of pressure $p$ versus  time, in terms of cycles N, (Fig. 5a) selected from
the examined time series (Fig. 1). The trajectory of succeeding  cycles 1,2,3,4 marked in Fig. 5a is
presented in Fig. 5b.
\end{enumerate}

\newpage

\begin{figure}

\begin{center}
\epsfig{file=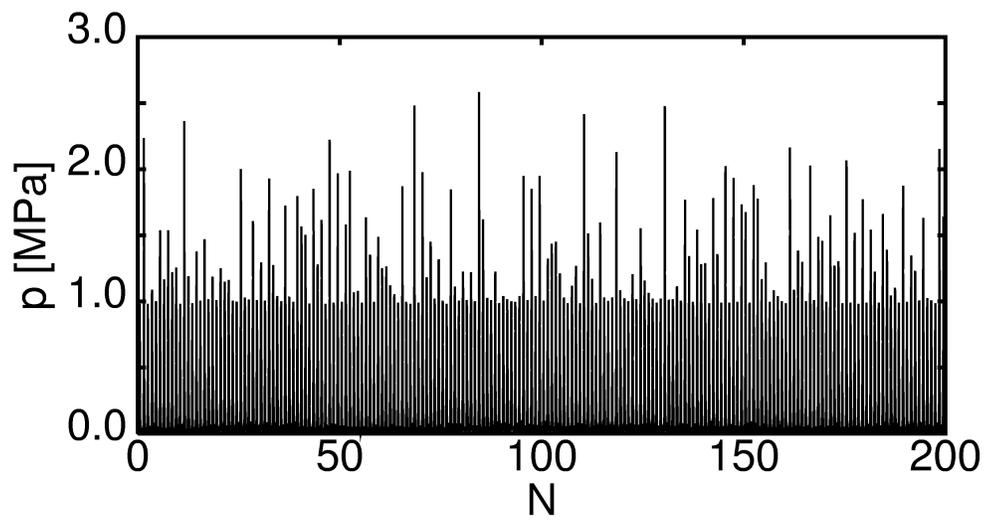,width=7.5cm,angle=-90}
\end{center}

\caption{\label{rys1}  Experimental time series of pressure $p(N)$ in a combustion process for
a crankshaft frequency $\Theta=600$ RPM and  a spark advance  angle $\Delta \alpha_z=\pi/6$.
Note, time is represented by sequential cycles $N$.
}
\end{figure}

\begin{figure}

\begin{center}
\epsfig{file=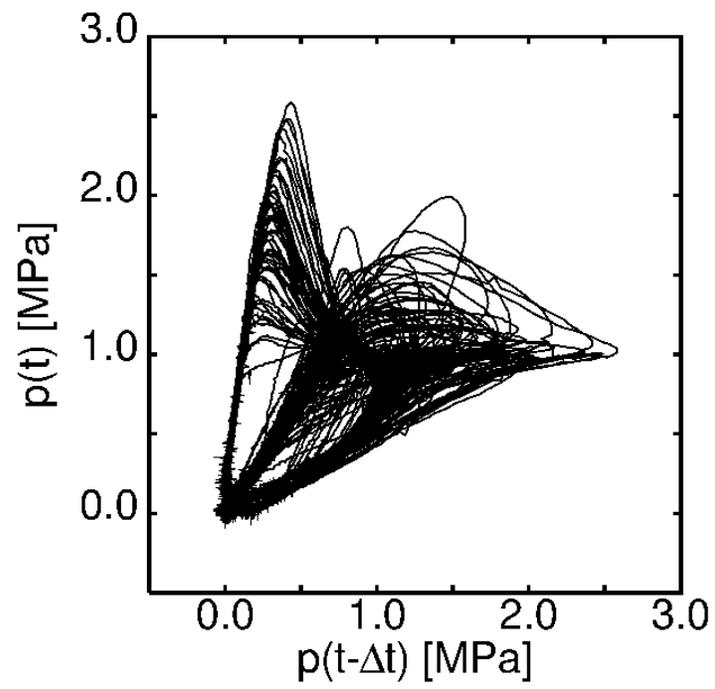,width=10.5cm,angle=-90}
\end{center}

\caption{\label{rys2} A 'butterfly' attractor reconstructed from  the experimental
time series of pressure. $\Delta t=0.977 T$.
}
\end{figure}

\begin{figure}

\begin{center}
\vspace{-2.5cm}
~

\vspace{-0.3cm}
\epsfig{file=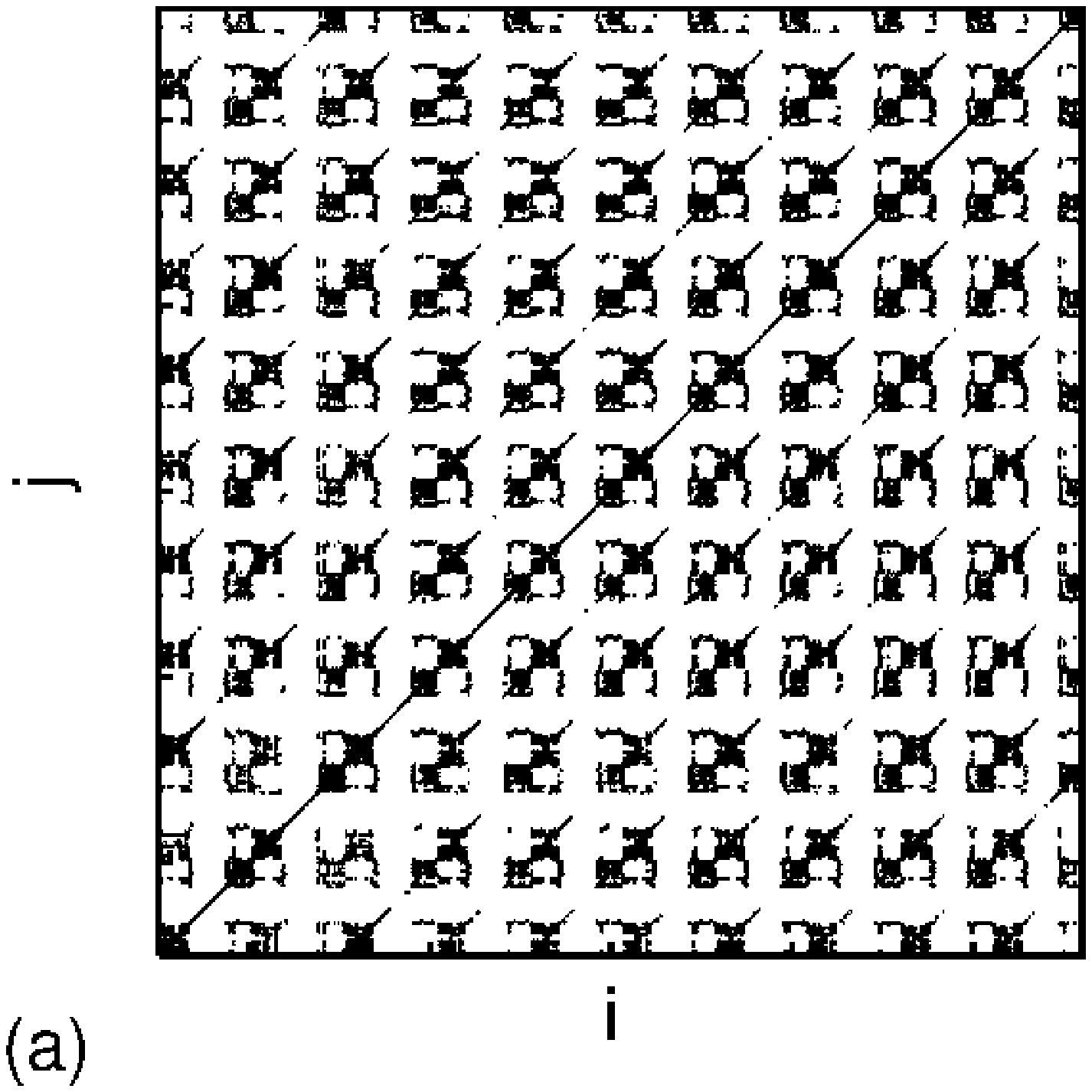,width=6.0cm,angle=-90}

\vspace{-0.3cm}
\epsfig{file=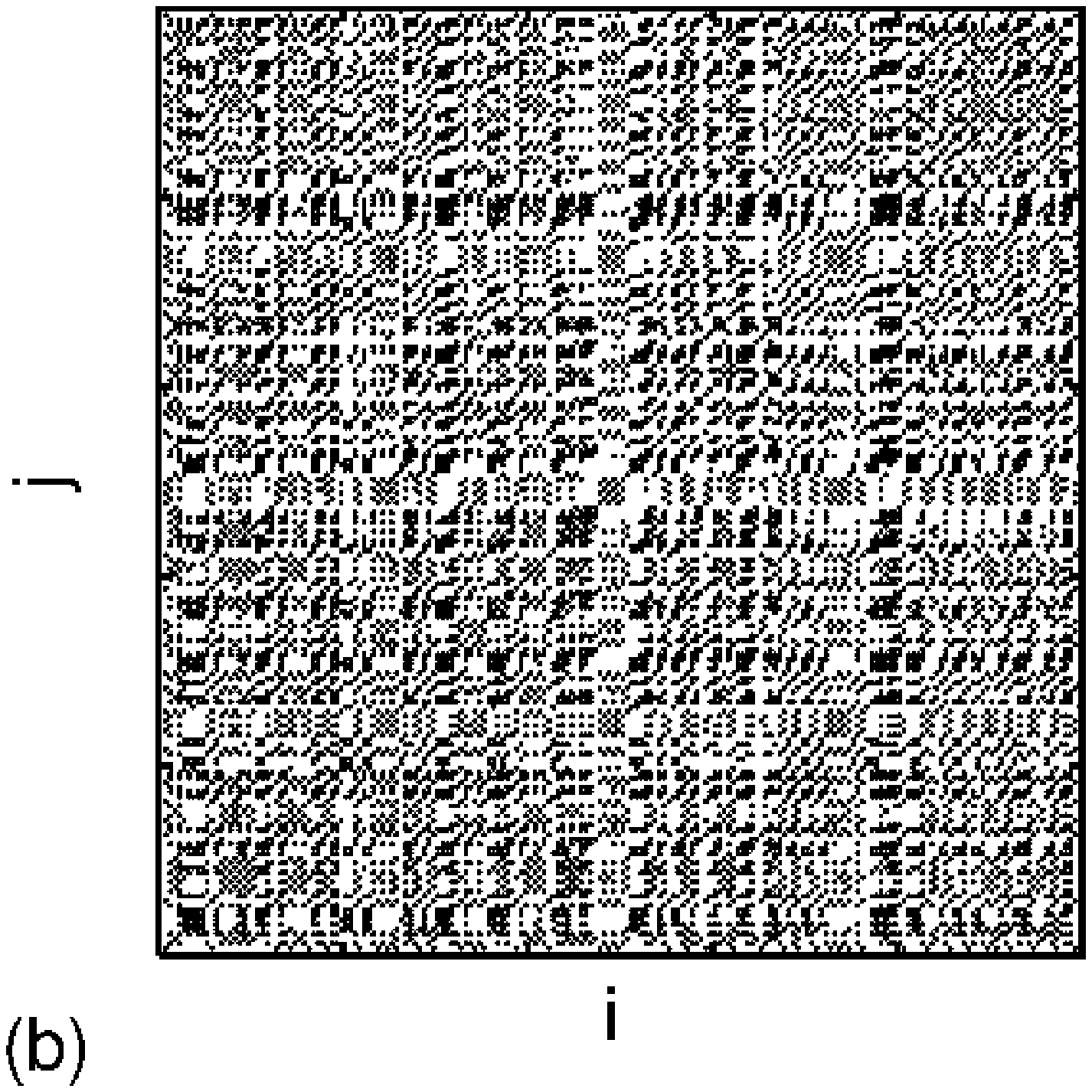,width=6.0cm,angle=-90}

\vspace{-0.3cm}
\epsfig{file=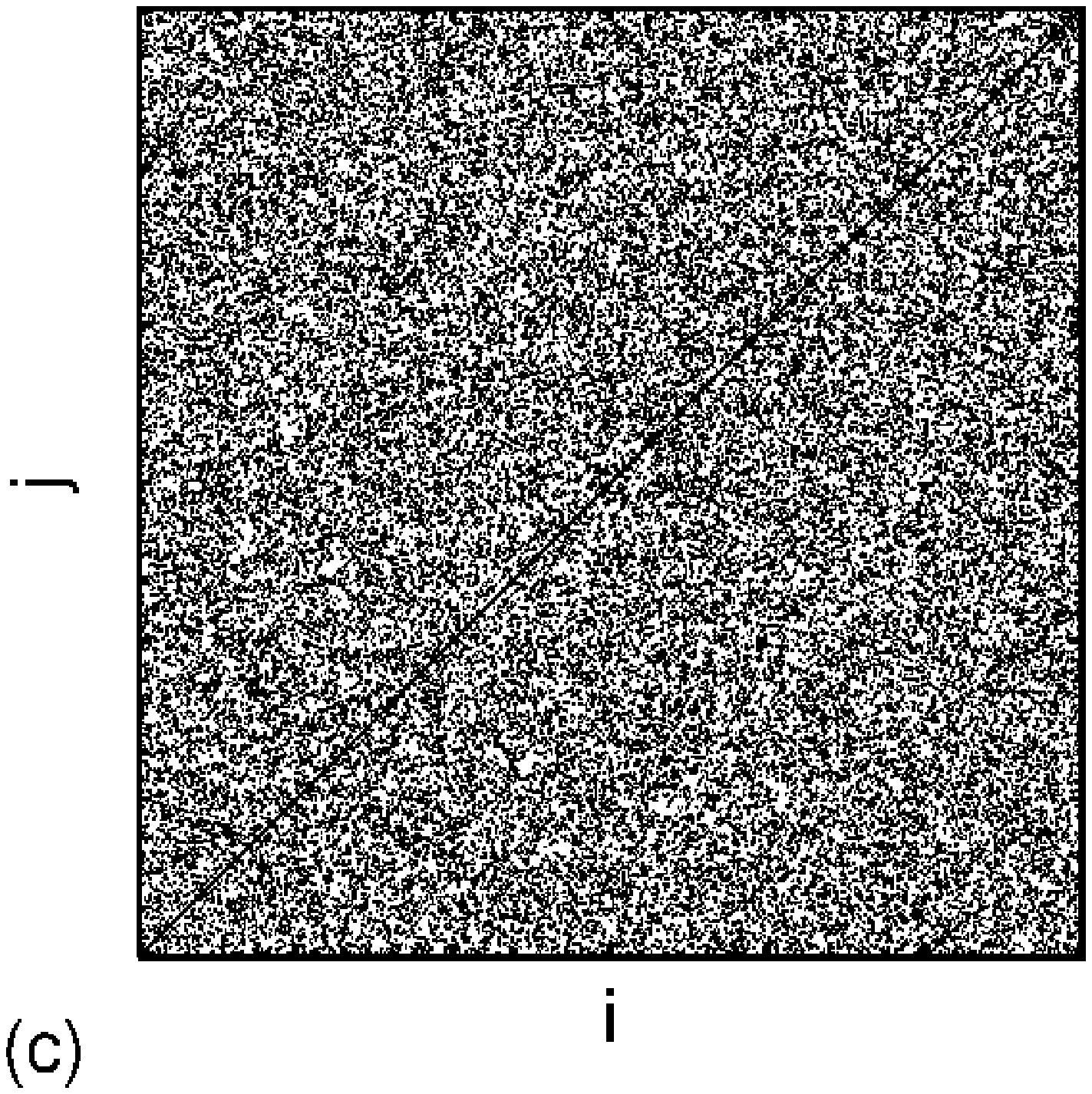,width=6.0cm,angle=-90}
\end{center}

\caption{\label{rys3}
Recurrence plots of the examined combustion process: in a small scale of 10 cycles (Fig.
3a),
and the scale  of 200 cycles for selected points chosen for the same crank angle $\Theta=2.125\pi$ (Fig. 
3b).
Fig.
3c shows the  recurrence plot for the series of numbers from random numbers generator with a uniform
distribution.
}
\end{figure}

\begin{figure}

\begin{center}
\vspace{-2.5cm}
~

\vspace{-0.2cm}
\epsfig{file=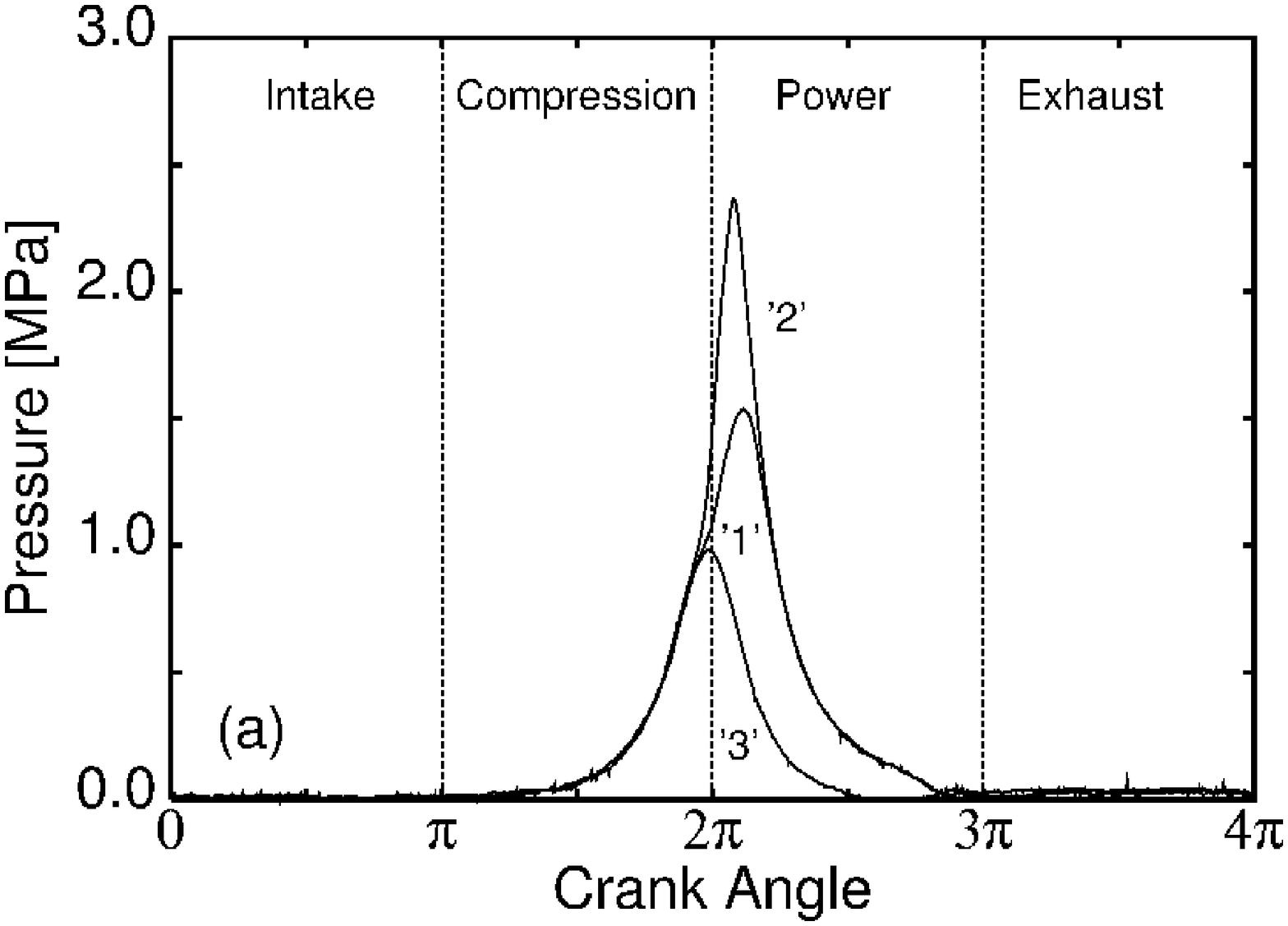,width=6.0cm,angle=-90}

\vspace{-0.2cm}
\epsfig{file=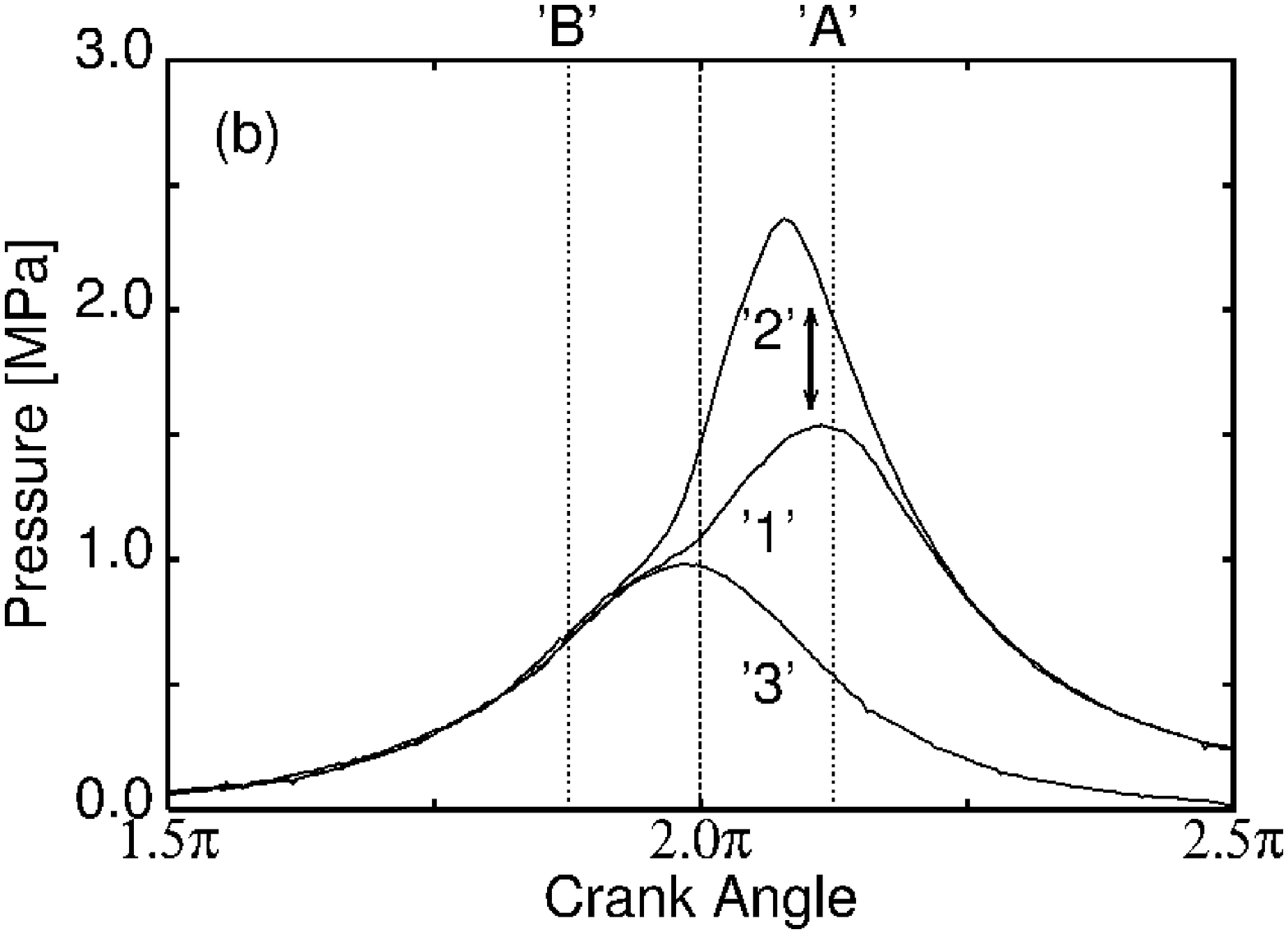,width=6.0cm,angle=-90}

\vspace{-0.2cm}
\epsfig{file=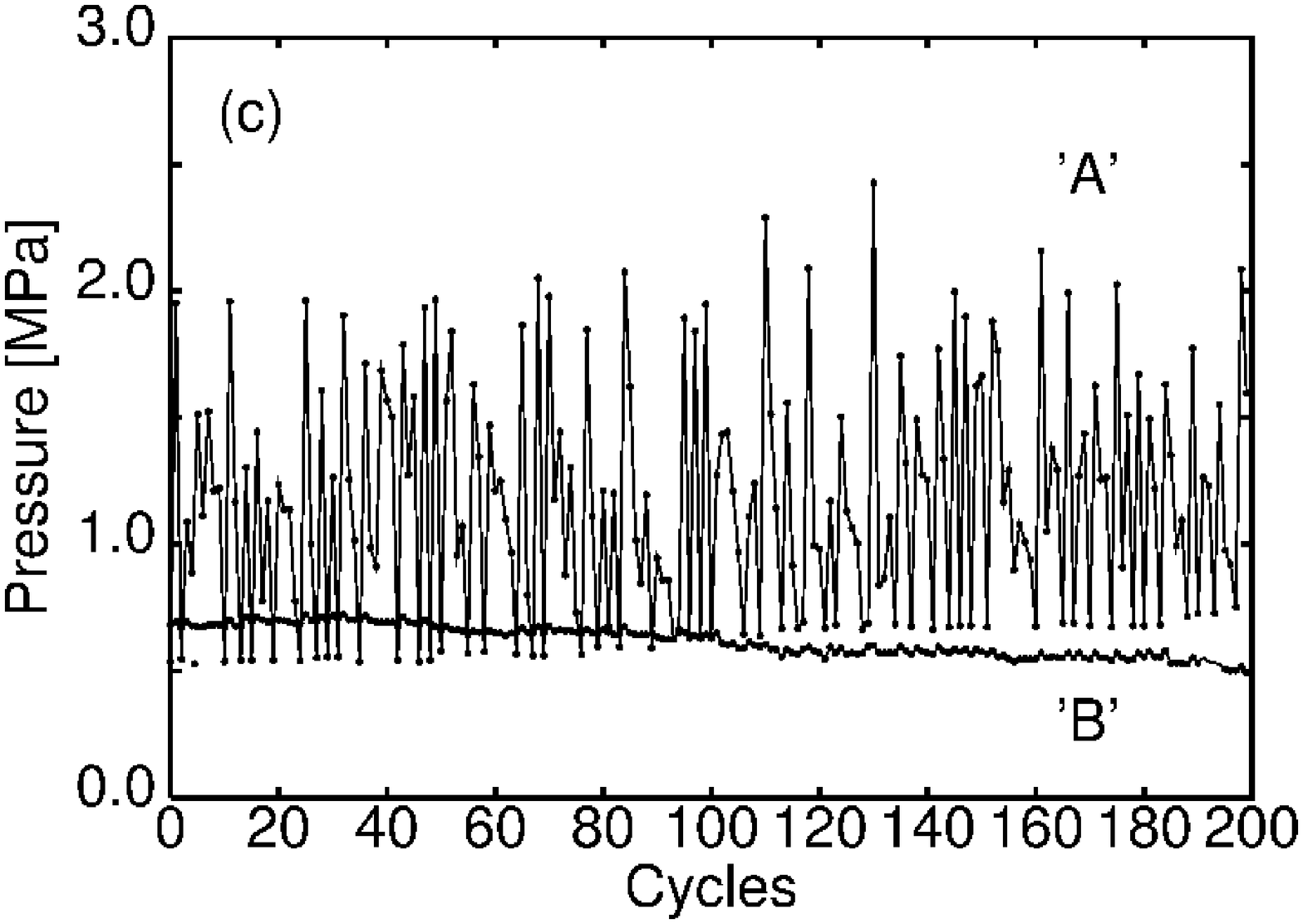,width=6.0cm,angle=-90}
\end{center}

\caption{\label{rys4} 
Time history  of combustion presented in Fig. 1 plotted for three chosen cycles (Fig. 4a)
and
magnification of it (Fig. 4b) around the top dead point (for a crank angle $\phi=2\pi$). Fig. 4c shows a 
sequence 
of
pressure
values for given crank angle $\phi=2\pi \mp 0.125 \pi$ (for B and A as before and after ignition,
respectively).
}
\end{figure}

\begin{figure}

\begin{center}
\vspace{-2.5cm}
~

\vspace{-0.7cm}
\epsfig{file=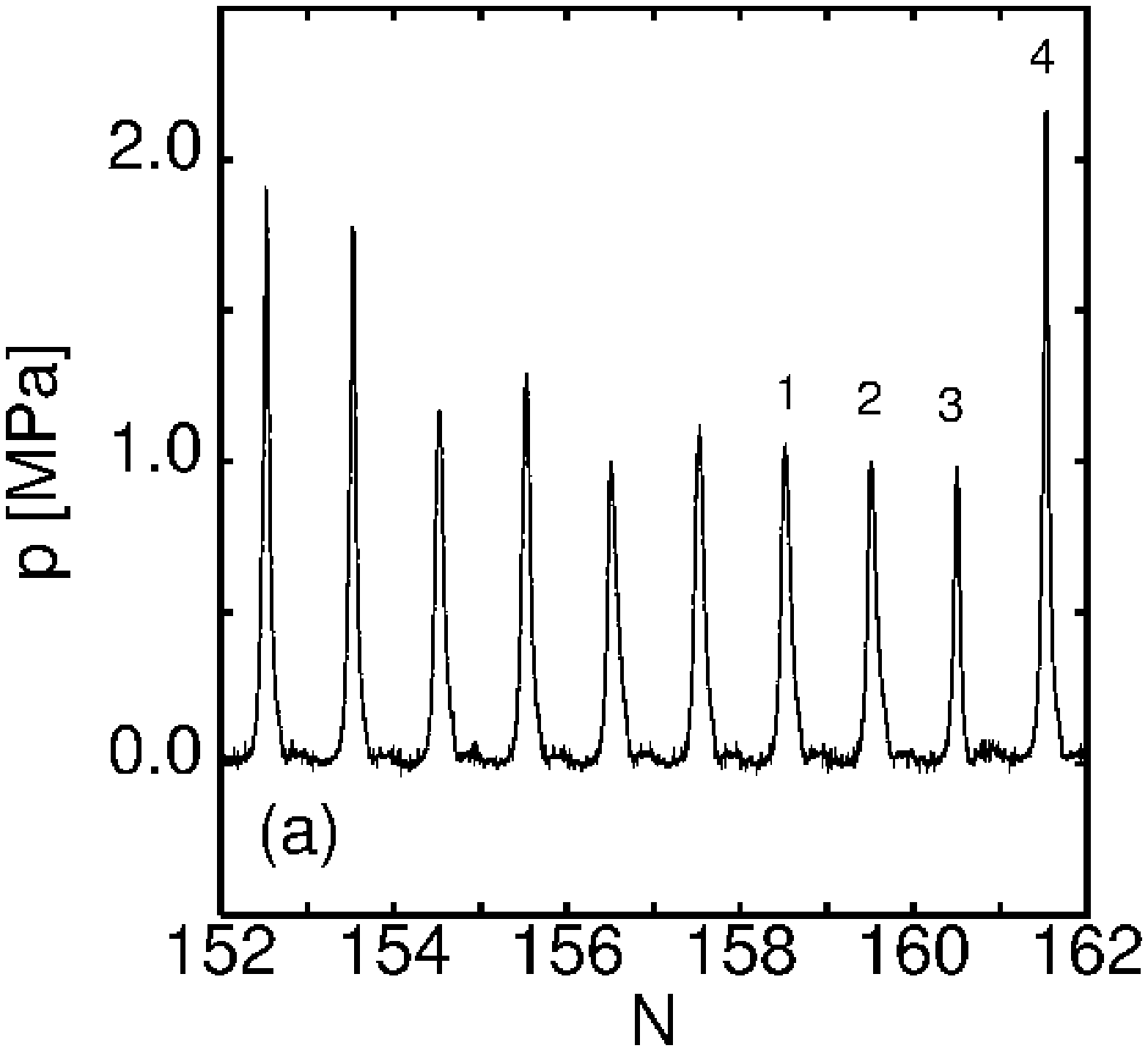,width=8.5cm,angle=-90}

\epsfig{file=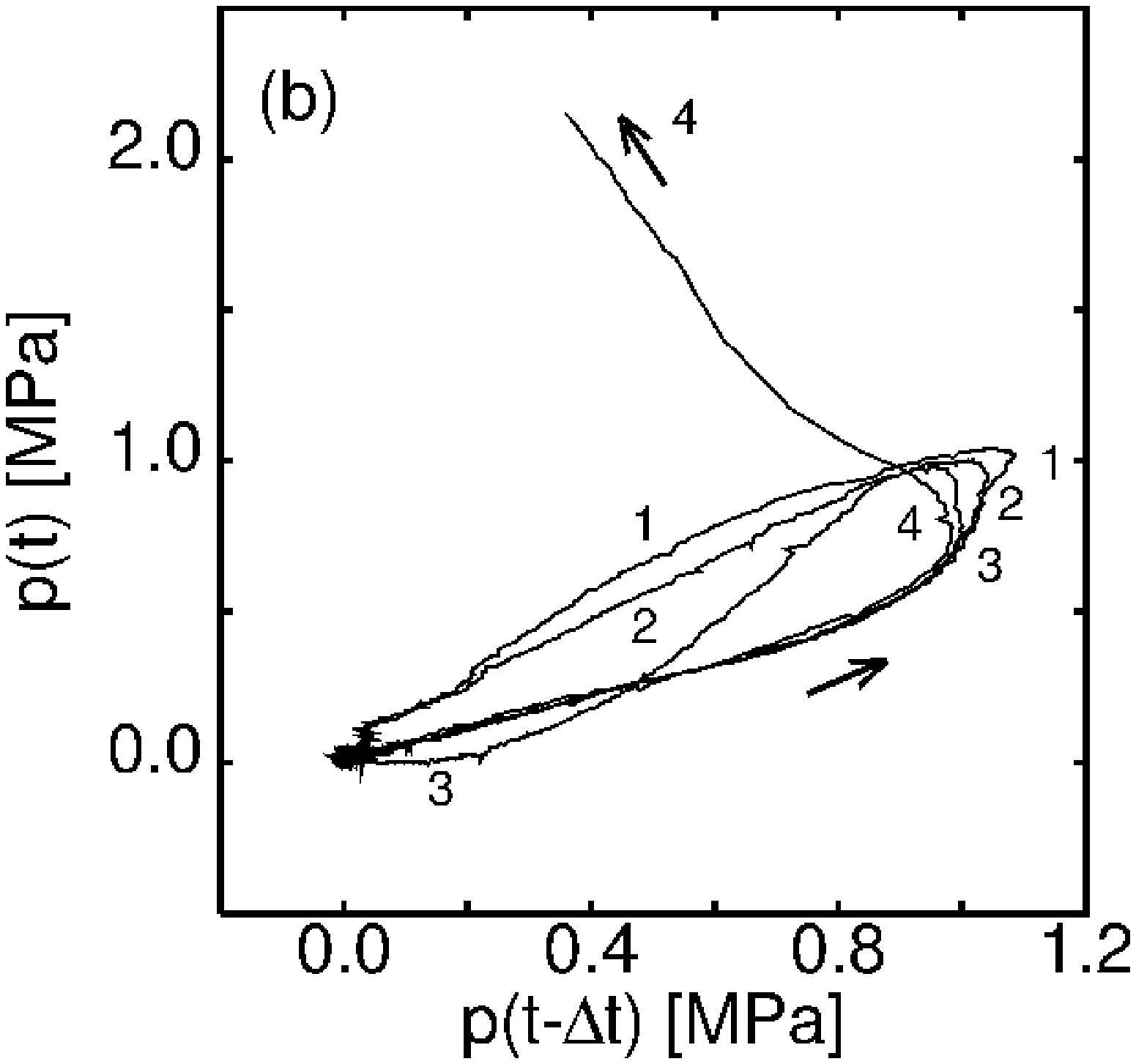,width=8.5cm,angle=-90}
\end{center}

\caption{\label{rys5}
Interval of the time history of pressure $p$ versus  time, in terms of cycles N, (Fig. 5a) selected from
the examined time series (Fig. 1). The trajectory of succeeding  cycles 1,2,3,4 marked in Fig. 5a is
presented in Fig. 5b.
}
\end{figure}

\end{document}